\documentclass[11pt]{article}

\usepackage{blois,epsfig}

\def\be{\begin{equation}}
\def\ee{\end{equation}}
\def\bea{\begin{eqnarray}}
\def\eea{\end{eqnarray}}

\usepackage{array}
\newcolumntype{L}[1]{>{\raggedright\let\newline\\\arraybackslash\hspace{0pt}}m{#1}}
\newcolumntype{C}[1]{>{\centering\let\newline\\\arraybackslash\hspace{0pt}}m{#1}}
\newcolumntype{R}[1]{>{\raggedleft\let\newline\\\arraybackslash\hspace{0pt}}m{#1}}

\newcommand{\Photo}{}

\begin{document}
\vspace*{4cm}
\title{SEARCHES FOR AXIONS WITH THE EDELWEISS EXPERIMENT}

\author{ Thibault DE BOISSI\`{E}RE\footnote{thibault.main-de-boissiere@cea.fr} (for the EDELWEISS collaboration) }

\address{CEA, Centre d'\'{E}tudes Saclay, IRFU, 91191 Gif-Sur-Yvette Cedex, France}
 
\maketitle\abstracts{The EDELWEISS experiment primarily aims at the direct detection of WIMPs using germanium bolometers. It is also sensitive to the low-energy electron recoils that would be induced by axions. We present new constraints on the couplings of axions using data from the EDELWEISS-II experiment. Using a total exposure of up to 448~kg.d, we searched for axion-induced electron recoils down to 2.5~keV within four scenarios involving different hypotheses on the origin and couplings of axions. We set a 95~\% CL limit on the coupling to photons $g_{A\gamma}<2.15\times 10^{-9}$~GeV$^{-1}$ in a mass range not fully covered by axion helioscopes. We constrain the coupling to electrons, $g_{Ae} < 2.59\times 10^{-11}$, similar to the more indirect solar neutrino bound. Finally we place a limit on $g_{Ae}\times g_{AN}^{\rm eff}<4.82 \times 10^{-17}$, where $g_{AN}^{\rm eff}$ is the effective axion-nucleon coupling for $^{57}$Fe. Combining these results we fully exclude the mass range $0.91\,{\rm eV}<m_A<80$~keV  for DFSZ axions and $5.73\,{\rm eV}<m_A<40$~keV for KSVZ axions.}

\section{The EDELWEISS experiment}

Cosmological observations indicate that 27\% of the universe is made of Dark Matter\cite{bib:Nakamura}. Despite decades of experimentations, its nature remains unknown. One attractive candidate (referred to as a Weakly Interacting Massive Particle or WIMP) arises from theories beyond the Standard Model like SUSY. Its mass and weak cross section naturally provide the observed relic density. Galactic WIMPs are expected to scatter off nuclei so direct detection experiments search for nuclear recoils in massive detectors. The experimental challenge (background rejection at $\sim$ keV energies) motivates the use of radio-pure materials in clean environments. The EDELWEISS experiment operates germanium bolometers at very low temperatures (18~mK) in the Underground Laboratory of Modane. They are protected from external radioactivity by lead and polyethylene shields and also an active muon veto. For each detector a set of interleaved electrodes and thermometers measure the ionization and phonon signals triggered by incoming particles. The comparison of the two signals allows a separation of nuclear recoils from electron recoils induced by $\beta$ and $\gamma$ radioactivity. These electron recoils constitute the major source of background in most direct WIMP searches. With the ionization signals, we also reject near-surface interactions by defining a fiducial volume for each detector. EDELWEISS-II has provided frontier sensitivities to WIMP-nucleon cross-sections for WIMP masses above 50 GeV~\cite{ref_EDW2first,ref_EDW2}, as well as for low-mass WIMPs $\sim 10$~GeV~\cite{lowmasspaper}. Since the EDELWEISS experiment is also sensitive to electron recoils, it can be used to detect new particles that interact with electrons. One such candidate, the axion, was postulated to solve the CP problem of QCD\cite{bib:pecceiandquinn}. While the original axion model was quickly dismissed by subsequent experiments, "invisible" axions with Standard Model interactions and the symmetry breaking scale as a free parameter are still viable. The purpose of this paper is to report on the axion searches results from the EDELWEISS experiment. More details can be found in~\cite{papieraxion}.

\section{Axion searches}

\subsection{Production and detection}\label{subsec:prod_detec}

Two classes of axion models stand out: DFSZ and KSVZ models. Despite their differences, both predict couplings to standard model particles. The Sun is expected to be a rich source of axions through various processes (see Fig.~1, left) which notably involve three couplings: $g_{Ae}$, $g_{A\gamma}$, $g_{AN}^{\rm eff}$, respectively the coupling to electrons, photons, and nucleons. In addition, we study the interpretation of the DAMA modulation as a possible detection of Dark Matter axions.  

~\\
Through their weak coupling either to photons or to electrons, axions generate electron recoils which can be detected in the EDELWEISS bolometers. We studied two detection channels. The Axio-electric effect ($\propto g_{Ae}^2$) is the equivalent of the photo-electric effect with the absorption of an axion instead of a photon: A+e$^-$+Z$\rightarrow$ e$^-$+Z. Through the Primakoff effect ($\propto g_{A\gamma}^2$), axions can be converted into photons in the intense electric field of the germanium crystal. In this case, we take advantage of Bragg diffraction to enhance the signal.~\cite{bib:primakoff_theory}.

~\\We studied four scenarios summarized in Table~\ref{table_channels}. In each scenario, the combination of production and detection mechanisms implies that the expected axion signal is related to a given coupling or product of couplings. In the absence of a signal, we derive limits on the couplings in a model-independent way but the results can be interpreted in the light of a particular model.

\begin{table}[!ht]
\begin{center}
\begin{tabular}{|C{7cm}|c|c|} 
  \hline
  Production &  Detection & Signal  \\
  \hline
 Primakoff in the Sun & Primakoff (Bragg scattering) & $\propto g_{A\gamma}^4$\\
  \hline
  $^{57}$ Fe magnetic transition in the Sun & Axio-electric & $\propto \left(g_{AN}^{\rm eff}g_{Ae}\right)^2$\\
  \hline
   Compton, Bremsstrahlung and recombination-deexcitation in the Sun  & Axio-electric & $\propto g_{Ae}^4$\\
  \hline
  Galactic Dark Matter axion & Axio-electric & $\propto g_{Ae}^2$\\
  \hline
\end{tabular}
\end{center}
\caption{List of all four scenarios under study in this paper.}
\label{table_channels}
\end{table}

\begin{figure}[!ht]
\label{spectrum_data}
\centering
\includegraphics[width=0.49\textwidth]{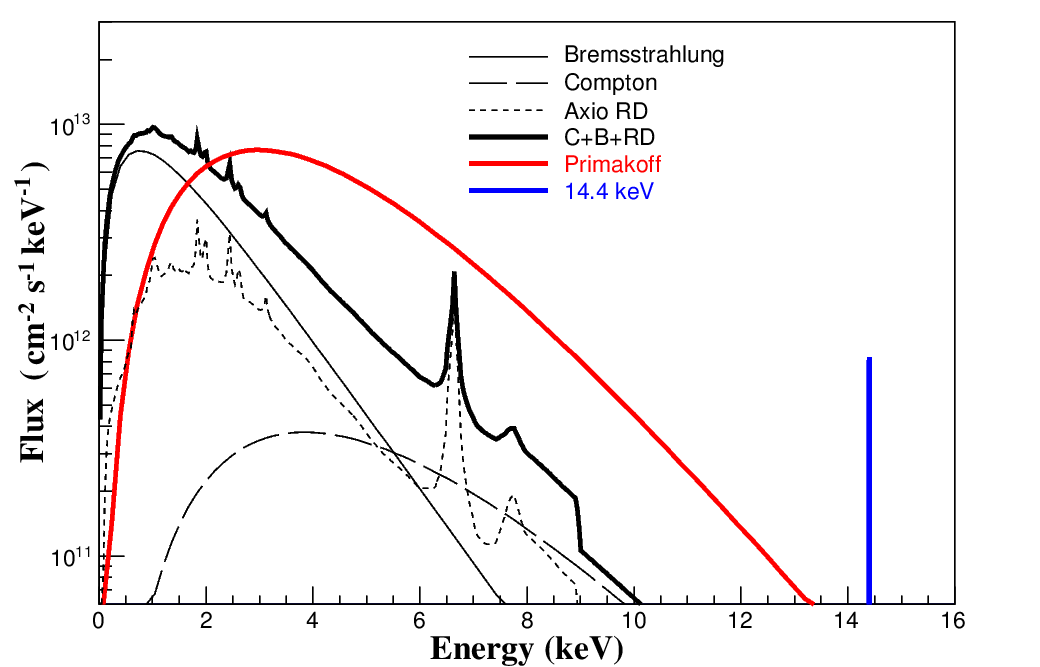}
\includegraphics[width=0.49\textwidth]{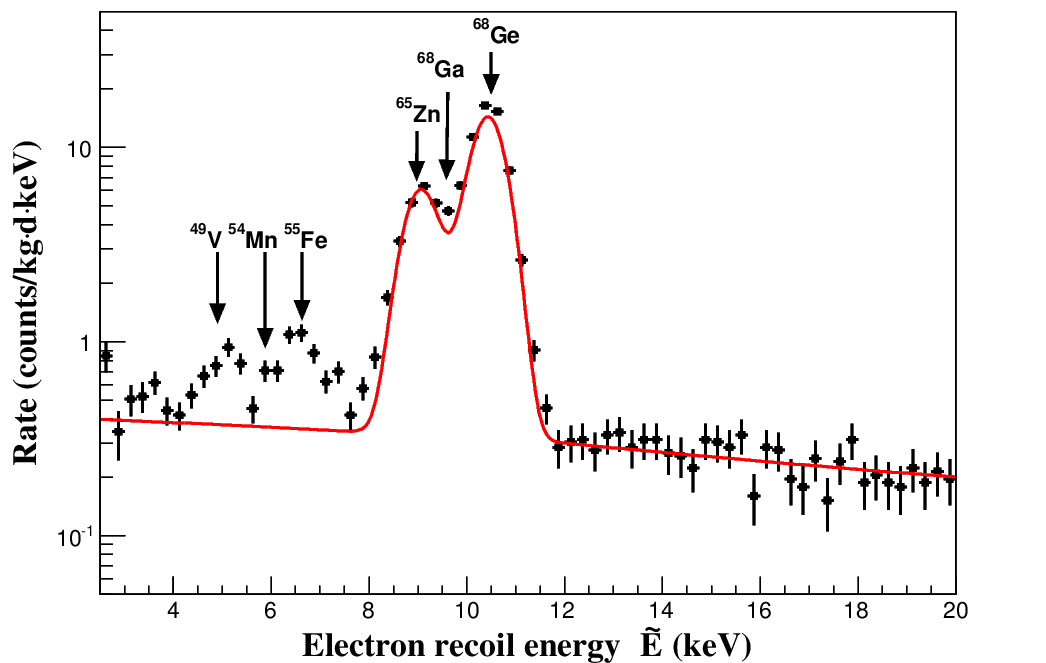}
\caption{Left: Predicted solar axion fluxes in the EDELWEISS detectors from different mechanisms.~Right: Stacked, efficiency-corrected electron recoil spectrum for the full exposure in the $2.5-18$~keV range. We also indicate identified cosmogenic X-ray lines. The red line is the conservative background model $B(\tilde{E})$ used in all analyses but Primakoff.}
\end{figure} 

\clearpage
\newpage

\subsection{Data analysis}\label{subsec:data_ana}

The data analysis is similar to the one performed for WIMPs~\cite{papieraxion}. We use data taken with ten 400-g bolometers in 2009-2010. The main difference is that data cuts were tailored to select the electron recoil band rather than the nuclear recoil band. Fiducial electron recoil events are selected by constraining the difference between the ionization and phonon energy. After this data selection, the observed background (see Fig.~1, right) consists of a Compton profile with a smooth, slightly decreasing energy dependence, together with radioactive peaks. Depending on the scenario, two statistical analyses were performed to look for axions.
\begin{enumerate}
  \item Axio-electric detection: We looked for a specific spectral structure in the electron recoil spectrum with a likelihood analysis.
  \item Primakoff detection: We used the time correlation of the count rate with the position of the Sun in the sky to improve the background rejection. We adapted the method in~\cite{Cebrian}. Monte Carlo simulations were performed to derive constraints on the signal.
\end{enumerate}

\subsection{Results}\label{subsec:results}

We found no evidence of an axion signal in any scenario. Therefore, we set model-independent constraints on the couplings of axions or axion-like particles to photons, electrons, and nucleons. Table~\ref{table_limits} summarizes the limits obtained for each channel on the respective couplings. Fig.~2 charts our limits on $g_{A\gamma}$ and $g_{Ae}$ as a function of the axion mass.

\begin{table}[!ht]
\begin{tabular}{|c|c|c|c|c|} 
  \hline
  Channel &  14.4 ($g_{Ae}\times g_{AN}^{\rm eff}$) & DM ($g_{Ae}$) & C-B-RD ($g_{Ae}$) & P ($g_{A\gamma}$)  \\
  \hline
 Limit  & $<4.82\times 10^{-17}$ & $<1.07\times 10^{-12}$ & $<2.59\times 10^{-11} $ & $<2.15\times 10^{-9}$~GeV$^{-1}$\\
  \hline
\end{tabular}
\caption{Summary of the limits on the different axion couplings. 14.4 stands for 14.4 keV solar axions, DM for dark matter axions, C-B-RD for Compton-bremsstrahlung and recombination-deexcitation axions, and P for Primakoff axions. The quoted values are in the limit $m_A\ll 1 {\rm keV}$, except for the dark matter case, which is given for $m_A=12.5$~keV. All limits are at 90\% CL except P (95\% CL).}
\label{table_limits}
\end{table}

\begin{figure}[!ht]
\centering
\includegraphics[width=0.48\textwidth]{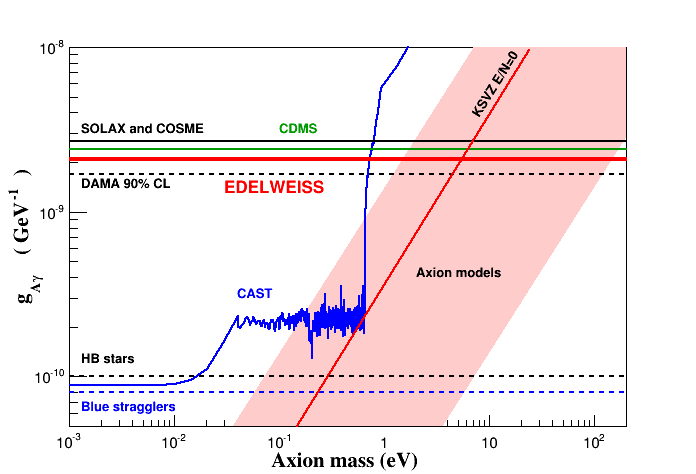}
\includegraphics[width=0.48\textwidth]{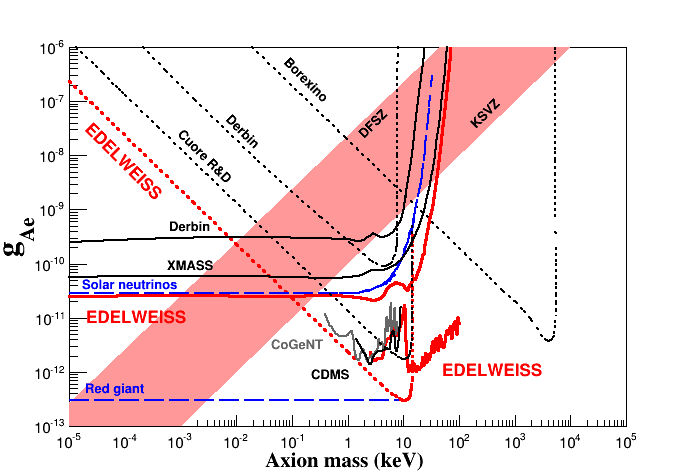}
\caption{Left: 95~\% CL limit on the $g_{A\gamma}$ coupling using the solar Primakoff flux obtained by EDELWEISS-II (red), compared to other crystal experiments. We also show the CAST limit and indirect bounds from HB stars and blue stragglers. The light red band labeled `Axion models' represents typical theoretical models. The red solid line inside this band represents the original KSVZ model.~Right: Constraints obtained by EDELWEISS-II on the $g_{Ae}$ axion coupling as a function of $m_A$. Continuous lines starting at $m_A=0$: model-independent limit on $g_{Ae}$ from the solar CBRD flux. Continuous lines in the keV mass range: limits on the coupling of  axions assuming they constitute all local galactic dark matter. Dotted lines: bounds derived from constraints on $g_{Ae}\times g_{AN}^{\rm eff}$ on various nuclei by assuming that $g_{AN}^{\rm eff}$ follows the DFSZ model with $\cos \beta_{\rm DFSZ}=1$. Benchmark DFSZ and KSVZ models are represented by a shaded band.~Astrophysical bounds: blue dashed lines.}
\label{limits}
\end{figure} 

\clearpage
\newpage

\section{Conclusion}

We analyzed data from the EDELWEISS-II detectors, originally used for WIMP searches, to constrain the couplings of axion-like particles within different scenarios. Contrary to WIMP searches, we used fiducial \textit{electron} recoils as a potential axion signal. In addition, the remarkable rejection of near-surface events provided by our detector design, primarily used to reject surface beta radioactivity in WIMP searches, also allows the rejection of low-energy gamma-rays of external origin. 

~\\We set new limits on axion parameters for different scenarios, some of which currently provide the best bounds for direct axion searches. The 95\% CL bound $g_{A\gamma}<2.15\times 10^ {-9}$~GeV$^{-1}$ derived from the solar Primakoff channel constrains axion models in the mass range $\sim 1- 100$~eV for KSVZ axions. This constraint is complementary to helioscope bounds, which can currently only probe lower axion masses within these models. Remarkably, the model-independent bound on $g_{Ae}$ obtained from the search for solar Compton-bremsstrahlung-RD axions reaches a better sensitivity than the more indirect bound derived from solar neutrino flux measurements.
Combining the results from all solar axion channels provides a wide model-dependent mass exclusion range, $0.91\,{\rm eV}<m_A<80$~keV within the DFSZ framework and $5.73\,{\rm eV}<m_A<40$~keV for KSVZ axions. This is a prominent result for a direct axion search from a single dataset.

~\\We therefore demonstrated the potential of germanium bolometric detectors for future axion-like particles searches. Improvements are expected with future experiments, such as EDELWEISS-III and EURECA~\cite{bib:EURECA}, thanks to both better energy resolution and larger exposures.

\end{document}